\documentclass[preprint,aps,pra,superscriptaddress,showpacs,floatfix]{revtex4-1}
\usepackage{graphicx}
\usepackage{xcolor}
\usepackage{amsmath}
\usepackage{amssymb}
\usepackage{amsfonts}
\usepackage{bm}
\usepackage[
   pdftitle={Scaling relations of the time-dependent Dirac equation 
describing multiphoton ionization of hydrogen-like ions}, 
   pdfauthor={I. V. Ivanova, V. M. Shabaev, Dmitry A. Telnov, A. Saenz},    
   pdfkeywords={time-dependent Dirac equation, highly charged ions, 
                multiphoton ionization},   
   colorlinks,
   linkcolor=blue,
   urlcolor=blue,
   citecolor=blue,
   ]{hyperref}
\usepackage{url}
\usepackage[english]{babel}
\usepackage[left=2cm,right=2cm,top=2cm,bottom=2cm]{geometry}
\graphicspath{{./figures/}}

\renewcommand{\vec}[1]{{\mbox{\boldmath$#1$}}}
\begin{document}
\title{
Scaling relations of the time-dependent Dirac equation describing 
multiphoton ionization of hydrogen-like ions }
\author{I.~V.~Ivanova}
\email{irina.ivanova@spbu.ru}
\affiliation{Department of Physics, St. Petersburg State University, 7-9 
Universitetskaya naberezhnaya, 199034 St. Petersburg, Russia}
\affiliation{NRC “Kurchatov Institute”, 1 Akademika Kurchatova pl., 123182 Moscow, Russia}
\author{V.~M.~Shabaev}
\affiliation{Department of Physics, St. Petersburg State University, 7-9 
Universitetskaya naberezhnaya, 199034 St. Petersburg, Russia}
\author{Dmitry~A.~Telnov}
\affiliation{Department of Physics, St. Petersburg State University, 7-9 
Universitetskaya naberezhnaya, 199034 St. Petersburg, Russia}
\author{Alejandro Saenz}
\affiliation{Institut f\"ur Physik,  Humboldt-Universit\"at zu Berlin,
Newtonstra{\ss}e 15, 12489 Berlin, Germany}
\begin{abstract}
Approximate scaling laws with respect to the nuclear charge are introduced 
for the time-dependent Dirac equation describing hydrogen-like ions 
subject to laser fields within the dipole approximation. In particular, 
scaling relations with respect to the laser wavelengths and peak intensities 
are discussed. The validity of the scaling relations is investigated for 
two-, three-, four-, and five-photon ionization of hydrogen-like ions with 
the nuclear charges ranging from $Z=1$ to $92$ by solving the corresponding 
time-dependent Dirac equations adopting the properly scaled laser parameters. 
Good agreement is found and thus the approximate scaling relations are 
shown to capture the dominant effect of the response of highly-charged 
ions to intense laser fields compared to the one of atomic hydrogen. 
On the other hand, the remaining differences are shown to allow for 
the identification and quantification of additional, purely relativistic 
effects in light-matter interaction.
\end{abstract}

%
\maketitle
\section{Introduction}
The development of light sources with extreme peak intensities remains an 
active field of research and technology. The Extreme Light Infrastructure (ELI)~\cite{eli1,eli2} strives for laser peak intensities of up to $10^{24}\,$W/cm$^2$ and free-electron lasers as the X-ray Free Electron Laser (XFEL)~\cite{xfel} at Hamburg and the Linear Coherent Light Source (LCLS)~\cite{lcls} at Stanford are expected to produce fields with peak intensities of up to $10^{25}\,$W/cm$^2$ and wavelengths down to 0.05\,nm. Especially in combination with mobile electron-beam ion traps (EBIT) these light sources can investigate the interaction of highly charged ions with extremely intense light. Moreover, the High-Intensity Laser 
Ion-Trap Experiment (HILITE) is under construction at GSI, Darmstadt. The goal 
of this experiment is to study the interaction of atoms and ions confined in a 
Penning trap and exposed to very intense laser light \cite{vogel, ringleb}. 
The intense laser field for the HILITE experiment will be provided by the 
Petawatt High-Energy Laser for Heavy Ion EXperiments (PHELIX). This facility 
produces intense laser fields with the peak intensities up 
to 10$^{21}$ W/cm$^2$~\cite{phelix}. It is planned to carry out experiments 
on ionization and excitation of highly charged ions (up to 
uranium) exposed to strong laser fields delivered by the PHELIX facility 
in the framework of the Stored Particles Atomic Physics Research Collaboration 
(SPARC) project. Clearly, a fully relativistic treatment of the 
ion-laser interaction will be required for the correct
theoretical description and interpretation of such experiments. 

These significant advances in the light-source technology have 
stimulated a considerable interest in the theoretical investigations of heavy 
one-electron ions exposed to electromagnetic radiation with extremely high 
frequencies and intensities. Many relativistic approaches for the description 
of the ion-laser interaction have been suggested 
recently~\cite{selsto,pindzola,pindzola2,vanne,rozenbaum,klaiber,klaiber2, 
ivanova,simonsen,ivanova2,kjellsson,kjellsson2,telnov}. They 
include simplified models based on the Coulomb-corrected relativistic 
strong-field approximation (SFA)~\cite{klaiber, klaiber2} as well as various 
full-dimensional solutions of the time-dependent Dirac equation (TDDE) 
\cite{selsto,pindzola,pindzola2,vanne,rozenbaum,ivanova,simonsen,ivanova2, 
kjellsson,kjellsson2,telnov}. Some studies ~\cite{pindzola,vanne,rozenbaum,ivanova} 
treat the interaction of the ion with the electromagnetic field within the 
so-called dipole approximation where the spatial dependence of the vector 
potential is neglected. The dipole approximation is a traditional approach for 
the infrared, visible, and ultraviolet light; in this frequency range it is 
usually well justified, since the wavelength exceeds by far the size of the ion. 
This is not necessarily the case for the hard X-ray radiation, and several 
attempts have been made to go beyond the dipole approximation taking into account 
the spatial properties of the laser 
pulse \cite{selsto,pindzola2,simonsen,ivanova2,kjellsson,kjellsson2,telnov}.  
If the photon energy and/or peak intensity of the laser pulse increase, the
non-dipole effects become more and more important, eventually making the 
theoretical description beyond the dipole approximation mandatory. However, 
for the experiments which will be carried out in the nearest future, a wide 
range of energies and intensities still exists where the dipole 
approximation is expected to be nevertheless reasonably well fulfilled.

Due to its relative simplicity, the hydrogen atom plays an important role 
in the understanding of light-matter interaction. In fact, the corresponding 
time-dependent Schr\"odinger equation (TDSE) can, at least within the dipole 
approximation, be solved efficiently for most of the practically relevant 
laser pulses. The results derived or obtained for atomic hydrogen may then 
be used to approximately predict the behavior of more complex atoms or 
even molecules in intense laser fields. For example, the 
Ammosov-Delone-Krainov (ADK) approximation \cite{adk} introduced effective 
quantum numbers that substitute the physical ones of hydrogen. In another 
approach, scaling relations were introduced in \cite{lambropoulos}. In fact, 
in \cite{madsen} it was shown that for hydrogen-like systems like positronium 
or highly-charged one-electron ions there exist \textit{exact} analytical 
scaling relations within the dipole approximation. In this case, the response 
of such systems to one laser pulse can be mapped onto the response of the 
hydrogen atom to a laser pulse with correspondingly scaled parameters. 

However, for the relativistic time-dependent Dirac equation describing 
atomic hydrogen exposed to an intense laser field, no \textit{exact} 
scaling relations could be found. Nonetheless, in \cite{vanne} an 
\textit{approximate} scaling law was proposed that matches the 
non-relativistic solutions of the TDSE and the relativistic solutions of 
the TDDE for a one-electron atomic ion. In the TDSE, an unphysical (scaled) 
nuclear charge is used that corrects the non-relativistic ionization potential 
to match the relativistic one. As is shown in \cite{vanne}, this gives 
good agreement between the TDSE results obtained with the scaled nuclear 
charge and the TDDE calculation with the physical nuclear charge in the 
considered multiphoton regime ranging from one- to five-photon ionization.   
In this work, the question of approximate scaling relations for the TDDE 
of hydrogen-like ions with respect to a variation of the nuclear charge 
is considered. Based on the scaling laws in \cite{madsen} and \cite{vanne}, 
we suggest a new, though approximate, scaling relation for the TDDE describing 
hydrogen-like ions exposed to intense laser fields. The validity of this 
scaling law is demonstrated by calculating the ionization yields 
of several hydrogen-like ions exposed to very short and intense laser pulses. 
For this purpose, we solve the TDDE numerically using the dipole approximation 
and length gauge. We report results adopting the TDDE scaling relations for 
ions with the nuclear charges ranging from 1 to 92 in the so-called multiphoton 
regime, considering variable laser wavelengths leading to two-, three-, four-, 
and five-photon ionization and a wide range of laser peak intensities 
(corresponding to a variation of two orders of magnitude).  

The approximate scaling relations in \cite{lambropoulos} were introduced 
in order to provide at least semi-quantitative predictions for the strong-field 
behavior of complex atoms based on theoretical results obtained for hydrogen or,  
especially experimental ones, for other atoms. Clearly, exact scaling relations  
allow even for quantitative predictions and evidently reduce the number of time-consuming TDSE calculations that need to be performed, as was discussed in \cite{madsen}. In fact, in the case of exact scaling relations 
experimentally found deviations would indicate experimental errors. Since the 
scaling relations in \cite{madsen} are derived within the non-relativistic 
Schr\"odinger theory and adopt the dipole approximation, experimentally found 
deviations would on the other hand indicate a break-down of either or both of these approximations. Depending on the experimental uncertainty, these deviations 
could even be quantified which is of great interest by itself. Since the magnitude 
of relativistic effects increases with the nuclear charge, a natural motivation 
for the present work is the question whether at least an approximate scaling 
relation can be found for the relativistic TDDE. This is of interest both 
for understanding the relevance and magnitude of various relativistic effects 
in light-matter interaction, but also in order to allow for at least some 
approximate prediction about the behavior of highly-charged one-electron 
atoms (or even more complex systems) in very intense laser fields. Comparing 
the TDDE results for one-electron ions with different nuclear charges the 
validity of the scaling relation (within the adopted dipole approximation) 
can be checked, see Secs.~\ref{ss:rel_scaling} and \ref{ss:intensity_scaling}. 
At the same time, the remaining deviations (that are truely 
of relativistic nature) can be quantified and their functional behavior can 
be analysed, as will be done in Sec.~\ref{ss:rel_effects}. In fact, scaling relations provide also a useful check for simplified models like, e.\,g., the ADK or the SFA theories.

For experimentalists scaling relations, though approximate, 
are helpful in planning experiments, since the laser parameters can 
be adjusted to, e.\,g., the efficiency for detecting the resulting fragments. 
This is otherwise a non-trivial problem, since in intense laser fields 
the ion yield increases easily by many orders of magnitude, if the intensity 
is varied. In fact, one of the main problems in experiments with light sources 
with very extreme peak intensities is the proper light-source characterization, 
including the determination of the peak intensity (see, for 
example,~\cite{alnaser,pullen,zhao,wallace}). On the basis of validated 
scaling relations highly-charged hydrogen-like ions may be used as 
a calibration tool, since one could compare the experimental results 
obtained when exposing an ion with a small nuclear charge to a 
well-characterised reference laser pulse that has a comparatively low 
intensity with the ones obtained for exposing a highly charged ion 
to the high-intensity laser pulse that should be characterised. 

The paper is organized as follows. In Sec.~\ref{s:theory} a simple scaling law of 
the TDDE is suggested and the method of solving the TDDE is described. The validation of the scaling relation for multiphoton ionization is 
discussed based on a wavelength scan covering the two- to five-photon 
regime in Sec.~\ref{ss:rel_scaling}. Sec.~\ref{ss:intensity_scaling} 
discusses the validity with respect to laser peak intensity. 
After the dominant relativistic effects have been shown to be considered by 
the scaling relations, Sec.~\ref{ss:rel_effects} investigates the remaining 
deviation not captured by the scaling relations. In Sec.~\ref{s:conclusion} 
the conclusions are given. Atomic units (a.u.) $\hbar = e = m_{e} = 1$ are 
used throughout this work unless otherwise specified. 
\section{Theory and computational method} \label{s:theory}
\subsection{Scaling of the TDDE with respect to the nuclear charge 
number} \label{ss:TDDE_scaling}
It is well-known that the TDSE for Coulomb systems interacting 
with external electromagnetic fields in the dipole approximation satisfies the
exact scaling laws with respect to the nuclear charge and the reduced mass 
(see \cite{madsen}). For example, the proper scaling of the spatial 
and the time variables in the equation itself as well as in the pulse parameters 
converts the TDSE for the hydrogen-like ion with the nuclear charge $Z$  into 
the TDSE for the hydrogen atom ($Z=1$). We shall refer to these scaling laws 
as the nonrelativistic scaling relations. They can be briefly summarized as %
\begin{equation} \label{eq:nonrel_sc}
\begin{split}
 r & \rightarrow r/Z, \\
 t & \rightarrow t/Z^2, \\
 \omega & \rightarrow \omega Z^2 \quad {\rm{(implying}} \: \lambda \: \rightarrow \: \lambda Z^{-2} {\rm{)}}, \\
 F_0 & \rightarrow F_0 Z^3 \quad {\rm{(implying}} \: I \: \rightarrow \: I Z^{6} {\rm{)}}, 
\end{split}
\end{equation}
where $r$ is the radial position coordinate, $t$ is the time, $\omega$ is the 
laser frequency, $\lambda$ is the wavelength, $F_0$ is the peak electric field 
strength, and $I$ is the laser peak intensity. 
If the dynamics of the hydrogen-like ion in the laser 
field is described by the TDDE, the same scaling laws do not apply, even if the 
dipole approximation is used. Discrepancies between the results of the
calculations with the original TDDE and that subject to the nonrelativistic 
scaling relations are shown and discussed below in Sec.~\ref{ss:nonrel_scaling}.

In general, deviations of the results obtained with the TDDE 
from the corresponding results obtained with the TDSE for the same system can be 
attributed to relativistic effects. In Ref.~\cite{vanne}, it was shown that 
the main relativistic effect is due to the shift of the ionization potential. 
A scaling relation was proposed to account for this effect (see Eq.~(27) in 
~\cite{vanne}). This relation suggests a scaled nuclear charge $Z'$ 
related to the true (physical) charge $Z$ {\it via} 
\begin{equation} \label{eq:scaled_charge}
  Z'= \sqrt{2c^2\left(1-\sqrt{1-Z^2/c^2}\right)}.
\end{equation}
As was shown for $Z=50$ in \cite{vanne}, 
calculations of multiphoton ionization using the TDSE with the scaled charge $Z'$ 
are in good agreement with the calculations using the TDDE and the true charge $Z$. 
At least, the scaling relation~\eqref{eq:scaled_charge} works well in the almost 
perturbative ionization regime considered in \cite{vanne}, confirming that the 
dominant relativistic effect in this case is the modification of the ionization potential. 
Other possible relativistic effects appeared to be negligibly small.  

In this work, we suggest a new \textit{approximate} scaling 
law for the TDDE describing hydrogen-like ions in laser fields. The approximate TDDE 
scaling relation implies that the behavior of the hydrogen-like ion with the nuclear 
charge $Z$ exposed to a laser pulse with the carrier wavelength $\lambda(=2\pi 
c/\omega)$ and peak intensity $I(=cF_0^2/8\pi)$ is \textit{almost} the same as 
that of the ion with the nuclear charge $\tilde{Z}$ exposed to a pulse with 
the carrier wavelength $\tilde{\lambda}$ and peak intensity $\tilde{I}$. Based 
on the previous results~\cite{vanne,madsen}, we derive the scaling relations 
between $\lambda$ and $\tilde{\lambda}$, $I$ and $\tilde{I}$ valid for a wide 
range of the nuclear charges. The principal idea is to combine the 
nonrelativistic scaling relation~\eqref{eq:nonrel_sc} with the scaling 
relation~\eqref{eq:scaled_charge}.

First, for any nuclear charges $Z$ and $\tilde{Z}$ we can 
calculate the scaled charges $Z'$ and $\tilde{Z'}$ from 
Eq.~\eqref{eq:scaled_charge}. Then, since the charges $Z'$ and $\tilde{Z'}$ 
represent the corresponding nonrelativistic systems described by the TDSE, the 
nonrelativistic scaling~\eqref{eq:nonrel_sc} can be used to obtain the 
relations %
\begin{equation}\label{eq:scaling}
\tilde{\lambda}=\lambda \left(\frac{Z'}{\tilde{Z}'}\right)^2; \ \ \ 
\ \tilde{I}=I\left(\frac{\tilde{Z}'}{Z'}\right)^6
\end{equation}
between the wavelengths and peak intensities. 
Finally, the scaling relations \eqref{eq:scaling} can be 
expressed through the true charges $Z$ and $\tilde{Z}$ with the help of 
Eq.~\eqref{eq:scaled_charge}, 
\begin{equation} \label{eq:scaling2}
\begin{split}
\tilde{\lambda} & =\lambda 
\frac{1-\sqrt{1-Z^2/c^2}}{1-\sqrt{1-\tilde{Z}^2/c^2}}; \\ 
\tilde{I} & =I\left(\frac{1-\sqrt{1-\tilde{Z}^2/c^2}}
{1-\sqrt{1-Z^2/c^2}}\right)^3.
\end{split}
\end{equation}
If the scaling relations~\eqref{eq:scaling2} are used for the laser 
parameters, the TDDE calculations for $Z$ and $\tilde{Z}$ are expected to be 
in good agreement with each other.  In the following, the method of solving 
the TDDE used in this paper is briefly introduced.
\subsection{Method of solving the TDDE} \label{ss:TDDE}
The relativistic dynamics of the hydrogen-like ion in the laser 
field is described by the TDDE 
\begin{equation} \label{eq:dirac}
 i\frac{\partial\Psi(t)}{\partial t}=H(t)\Psi(t),
\end{equation}
where $\Psi(t)$ is the time-dependent wave function of the electron, and the 
total Hamiltonian $H(t)$ can be represented as a sum of two terms,
\begin{equation} \label{eq:h_sum}
 H(t)=H_0+V(t).
\end{equation}
Here $H_0$ is the time-independent field-free Dirac Hamiltonian 
\begin{equation} \label{eq:field-free_H}
H_0=c(\vec\alpha \cdot \vec p)+c^2\beta+V_\mathrm{C}, 
\end{equation}
where $\vec{\alpha}$ and $\beta$ are the Dirac matrices.
We adopt the point-like nucleus model, thus the interaction $V_\mathrm{C}$ of 
the electron with the nucleus of the charge $Z$  is described by the Coulomb 
potential:
\begin{equation} \label{eq:coulomb}
V_\mathrm{C}=-\frac{Z}{r}.
\end{equation}
The interaction with the external electromagnetic field $V(t)$ is 
represented within the dipole approximation,
\begin{equation} \label{eq:length_gauge}
V(t)=\vec{r}\cdot\vec{F}(t)=zF(t),
\end{equation}
where $\vec{F}(t)=-d\vec{A}(t)/dt$ is the electric field 
strength, and $\vec{A}(t)$ is the vector potential. The 
field $\vec{F}(t)$ is assumed to be linearly polarized along the $z$ axis. In 
this work, we make use of the ion-laser interaction term in the length gauge; 
earlier it was shown that 
the observables obtained by solving the TDDE in the length and velocity gauges 
coincide with each other if the numerical convergence is 
reached~\cite{vanne,ivanova}.

Our scheme to solve the TDDE generally follows the approach 
described in Ref.~\cite{vanne}. At the first step, we solve the 
time-independent Dirac equation for the unperturbed (field-free) hydrogen-like 
ion where the electron moves in the Coulomb potential of the nucleus only. The 
field-free eigenstates can be found by either direct expansion of the
radial wave functions on a \textit{B}-spline~\cite{deboor} basis set
(see, for example,~\cite{johnson}) or with the help of the
dual-kinetic-balance (DKB) approach~\cite{shabaev}. Then the TDDE for the ion in 
the laser field is solved. The time-dependent Dirac wave function is expanded 
on the basis of the field-free eigenstates. The expansion coefficients can be 
found employing various propagation schemes. For example, we consider 
Crank-Nicolson propagation scheme~\cite{crank}, split-operator 
technique~\cite{fleck}, and variable-order, variable-step Adams 
solver~\cite{press}. Below we give a more detailed 
description of the algorithm for solving the TDDE.

To solve Eq.~\eqref{eq:dirac}, we expand the time-dependent Dirac wave function 
$\Psi(t)$ in a finite basis set which is represented by the eigenfunctions 
$\varphi_{n\kappa \mu}(\vec{r})$ of the field-free Hamiltonian,  
\begin{equation} \label{eq:expansion}
\Psi(\vec{r},t)=\sum_{n,\kappa}C_{n\kappa \mu}(t)e^{-iE_{n\kappa}t}\varphi_{n\kappa \mu}(\vec{r}),
\end{equation}
where $C_{n\kappa \mu}(t)$ are the expansion coefficients; the indices $n, 
\kappa$ define the full set of basis states, $n$ is the 
principal quantum number, $\kappa$ is the 
angular momentum-parity quantum number, and $\mu$ is the projection of the 
total electron angular momentum on the $z$ axis. The quantum number $\mu$ is 
conserved due to the axial symmetry along the $z$ axis. 

The angular momentum-parity quantum number $\kappa$ is 
expressed through the orbital angular momentum $l$ and total angular 
momentum $j$:
\begin{equation} \label{eq:kappa}
\kappa = (-1)^{l + j +1/2}(j+1/2).
\end{equation}
The time-independent and orthonormal basis functions 
$\varphi_{n\kappa \mu}(\vec{r})$ are the eigenfunctions of the unperturbed 
Hamiltonian $H_0$:
\begin{equation} \label{eq:h0_equation}
H_0\varphi_{n\kappa \mu}(\vec{r})=E_{n\kappa}\varphi_{n\kappa \mu}(\vec{r}),
\end{equation}
\begin{equation} \label{eq:stat_wf}
\varphi_{n\kappa \mu}(\vec{r})=\frac{1}{r}\left(\begin{tabular}{cc}$G_{n\kappa}(r)\Omega_{\kappa \mu}(\vec{n})$\\$iF_{n\kappa}(r)\Omega_{-\kappa \mu}(\vec{n})$\\\end{tabular}\right),\ \ \vec{n}=\frac{\vec{r}}{r},
\end{equation} 
where $G_{n\kappa}(r)$ and $F_{n\kappa}(r)$ are the upper and lower 
radial components of the wave function $\varphi_{n\kappa 
\mu}(\vec{r})$ while $\Omega_{\kappa \mu}(\vec{n})$ is the spherical spinor. The 
radial components can be calculated numerically by solving
ordinary differential equations. If $B$-spline expansions are straightforwardly 
used for this purpose (see, for example, Eq.~(13) in Ref.~\cite{vanne} or 
Eq.~(14) in Ref.~\cite{johnson}), then nonphysical (so-called spurious) states 
emerge among the solutions. To avoid such an undesirable effect, an appropriate 
modification of the $B$-spline basis set was suggested (the DKB approach~\cite{shabaev}). For the hydrogen-like ions, however, it 
is easy to identify and remove the spurious states even if the DKB approach is 
not used. Therefore in our case we can use both DKB and non-DKB schemes and 
achieve the same results.

By substitution of the expansion~\eqref{eq:expansion} into the 
TDDE~\eqref{eq:dirac}, the latter can be reduced to 
a set of first-order ordinary differential equations for 
the expansion coefficients,
\begin{equation} \label{eq:coeff}
i\frac{\partial}{\partial t}C_{K'}(t)= \sum_{K}V_{K'K}(t)C_{K}(t)e^{-i(E_{K}-E_{K'})t},
\end{equation} 
where the indices $K'$ and $K$ represent the full set of quantum numbers 
$\{n',\kappa',\mu\}$ and $\{n,\kappa,\mu\}$, respectively, 
and $V_{K'K}(t)$ is the time-dependent matrix element defined as
\begin{equation} \label{eq:matrix_element_short}
V_{K'K}(t)=\left<\varphi_{K'}\left|V(t)\right|\varphi_{K}\right>.
\end{equation}
In the length gauge, $V_{K'K}(t)$ may be written as
\begin{equation} \label{eq:matrix_element_long}
\begin{split}
V_{K'K}(t) & = F(t)(-1)^{j'+j+\frac{1}{2}-\mu}\sqrt{(2j'+1) (2j+1)} \\ 
& \times \int \limits_{0}^\infty{dr\; r \left[G_{n'j'l'}(r)G_{njl}(r)+F_{n'j'l'}(r)F_{njl}(r)\right]} \\
& \times \, \delta_{\left|l'-l\right|,1}\left(\begin{tabular}{ccc}$j'$&$1$&$j$\\$-\mu$&$0$&$\mu$
\end{tabular}\right)\left(\begin{tabular}{ccc}$j'$&$1$&$j$\\$-\frac{1}{2}$&$0$&$\frac{1}{2}$
\end{tabular}\right).
\end{split}
\end{equation}
The radial integration in the matrix elements~\eqref{eq:matrix_element_long} is 
performed numerically using the Gauss-Legendre 
quadrature, and the $3j$-symbol analytical expressions are 
obtained for the angular integrals~\cite{varshalovich}. 
With the matrix elements $V_{K'K}(t)$ at hand, the 
time propagation in Eq.~\eqref{eq:coeff} is carried out numerically.

In all the calculations reported here, the ground $1s_{1/2}$ 
electron state is chosen as the initial state for the time propagation. The 
projection $\mu$ of the total electron angular momentum is equal to $1/2$. We 
choose the same \textit{B}-spline basis set as in Ref.~\cite{vanne}, with 
500 \textit{B}-splines of the 9$^{\mathrm{th}}$ order. This number of 
\textit{B}-splines provides sufficient number of the continuum (both 
positive-energy and negative-energy) as well as bound states for each 
angular momentum. The radial box size $R = (250/Z)$~a.u. is adopted, as was 
suggested in Ref.~\cite{vanne} and can be understood from Eq.~\eqref{eq:nonrel_sc}.

The laser field is linearly polarized along the $z$ axis, 
and the vector potential is chosen in the form of a $N$-cycle cos$^2$-shaped 
pulse:
\begin{equation}\label{eq:pulse}
\vec{A}(t)= \left\{\begin{tabular}{cc}$\vec{e_z}A_0 \cos^2\left(\dfrac{\pi 
t}{T}\right)\sin(\omega t), $&$ \left|t\right|< T/2, $\\$0, 
$&$\left|t\right|\geqslant T/2, $\end{tabular}\right.
\end{equation}
where $\omega$ is the photon energy, $T$ is the pulse duration, $T=\frac{2\pi 
N}{\omega}$, and $A_0=F_0/ \omega$, $F_0$ is the peak electric field. We 
use the same laser pulse shape with $N=20$ in all our calculations.

After the calculation of all the expansion coefficients 
$C_{K}(t)$ on the time grid, the ionization probability 
can be found as a projection of the final electron wave function onto the 
states $\varphi_{K}$ with the energies higher than $mc^2$:
\begin{equation} \label{eq:ionization}
\begin{split}
P_{\mathrm{ion}}& = \sum_{\substack{K, \\ E_{K}\ge mc^2}
}\left|\left<\Psi(t=T/2)\left|\right.\varphi_{K}\right>\right|^2 \\ 
& = \sum_{\substack{K, \\ E_{K}\ge mc^2}}\left|C_{K}(t=T/2)\right|^2. 
\end{split}
\end{equation} 
%
\section{Results and discussion}  
\subsection{Scaling of the TDDE using the nonrelativistic scaling relations} \label{ss:nonrel_scaling}

\begin{figure}[tb]
\includegraphics[width=0.98\linewidth]{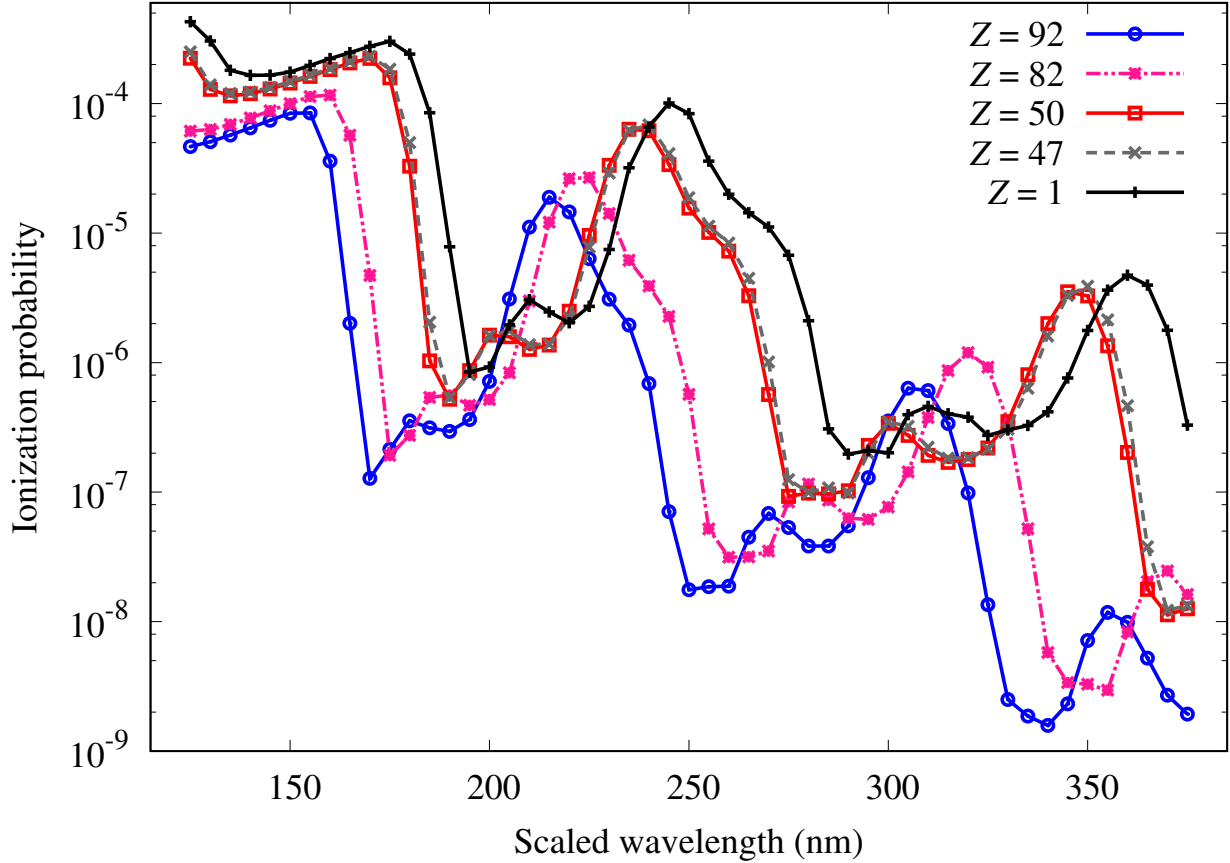}
\caption{Multiphoton ionization probability of the hydrogen-like ions with the 
nuclear charges $Z$ specified in the figure as a function of the scaled carrier 
wavelength $\lambda Z^{2}$. The laser pulse is $\cos^2$-shaped and contains 20 
optical cycles at each scaled wavelength. The peak intensity is equal to $5 
\times 10^{22}$~W/cm$^2$ for $Z=50$ and scaled according to 
Eq.~\eqref{eq:nonrel_sc} with $Z^6$ for the other nuclear charges.}
\label{fig:Z2_TDDE_scaling}
\end{figure}

First, we present the results of solving the TDDE for the 
hydrogen-like ions and laser pulse parameters after adopting the nonrelativistic 
scaling relations \eqref{eq:nonrel_sc}. Fig.~\ref{fig:Z2_TDDE_scaling} 
shows the multiphoton ionization probabilities of several hydrogen-like ions. 
The field parameters are the same as in Ref.~\cite{vanne} for $Z=50$ and 
properly scaled for the other nuclear charges. Our results for the ionization 
probability of the ion with the nuclear charge $Z=50$ are in good 
agreement with those presented in Fig.~5 of Ref.~\cite{vanne}.

Looking at the curves in Fig.~\ref{fig:Z2_TDDE_scaling}, 
one can see that the nonrelativistic scaling does not work satisfactorily for 
the TDDE, while it is exact for the TDSE. The curve for $Z=1$ essentially represents the 
nonrelativistic ionization probability because the relativistic effects are 
negligible for the hydrogen atom at the intensity and wavelengths used in the 
calculations (as tested by comparing the TDDE and TDSE results). Consequently, 
this curve also displays the ionization probabilities of the other hydrogen-like 
ions obtained by the nonrelativistic scaling. However, the curves corresponding to 
the higher nuclear charges and obtained by solving the TDDE are shifted from the 
curve for $Z=1$. The shifts can be explained as relativistic effects that become 
significant for highly charged ions and increase with the nuclear charge $Z$. 
For a narrow range of the $Z$ numbers, the nonrelativistic scaling 
approximately works even for the TDDE (see, for example, the results for 
$Z=47$ and $Z=50$ in Fig.~\ref{fig:Z2_TDDE_scaling}). However, for a wide $Z$ range, the 
nonrelativistic scaling does not work even approximately. A failure of the 
nonrelativistic scaling for the heavy hydrogen-like ions motivated us 
to search for different scaling relations that would work better with the TDDE. 
As a result, the scaling relation~\eqref{eq:scaling2} was derived.

\subsection{Scaling of the TDDE by the new scaling relations} \label{ss:rel_scaling}
%
\begin{figure}[tb]
\includegraphics[width=0.98\linewidth]{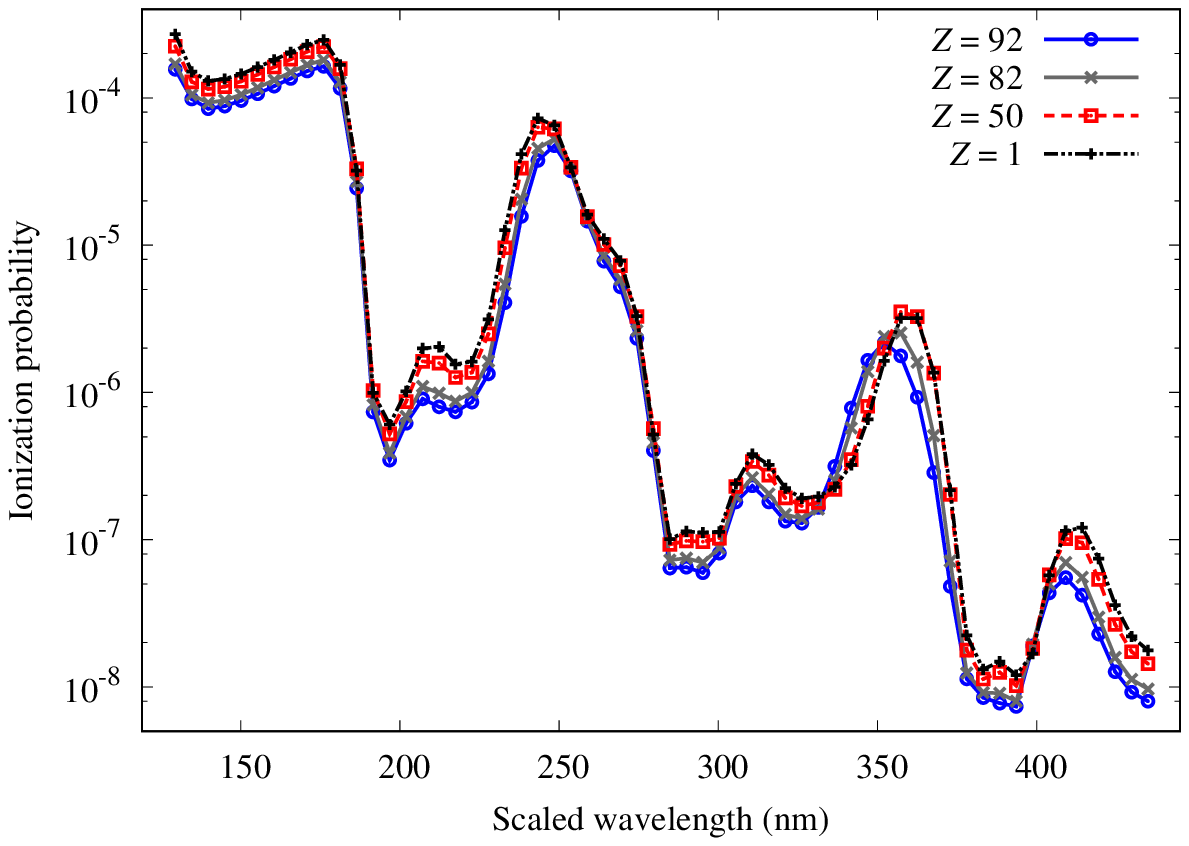}
\caption{Multiphoton ionization probability of the hydrogen-like ions with the 
nuclear charges $Z$ (specified in the figure) as a function of the scaled carrier 
wavelength $\lambda Z'^{2}$. $Z'$ is related to the true nuclear charge $Z$ by
Eq.~\eqref{eq:scaled_charge}. The laser pulse is $\cos^2$-shaped and contains 
20 optical cycles at each scaled wavelength. The peak intensity is equal to $5 
\times 10^{22}$~W/cm$^2$ for $Z=50$ and scaled according to 
Eq.~\eqref{eq:scaling2} for the other nuclear charges.}
\label{fig:New_TDDE_scaling}
\end{figure}
\begin{figure*}[tb]
\includegraphics[width=0.49\linewidth]{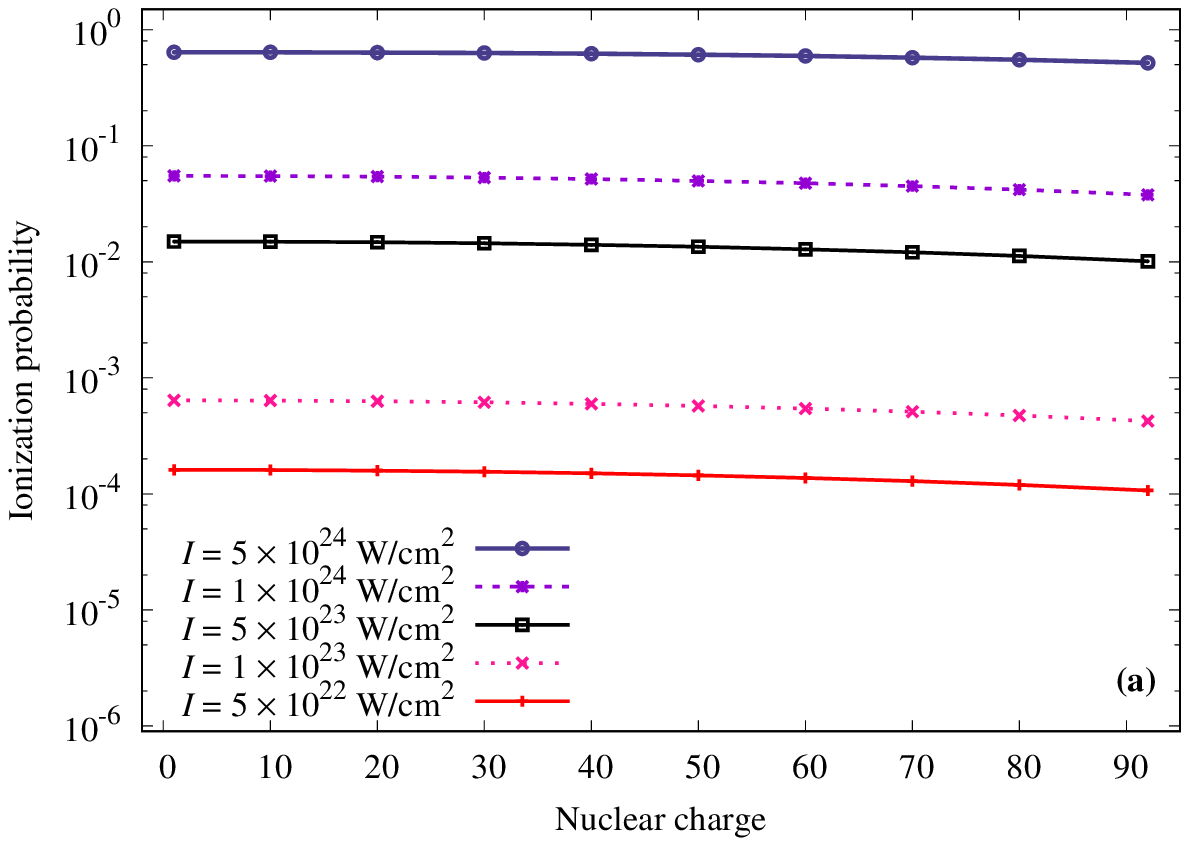}
\includegraphics[width=0.49\linewidth]{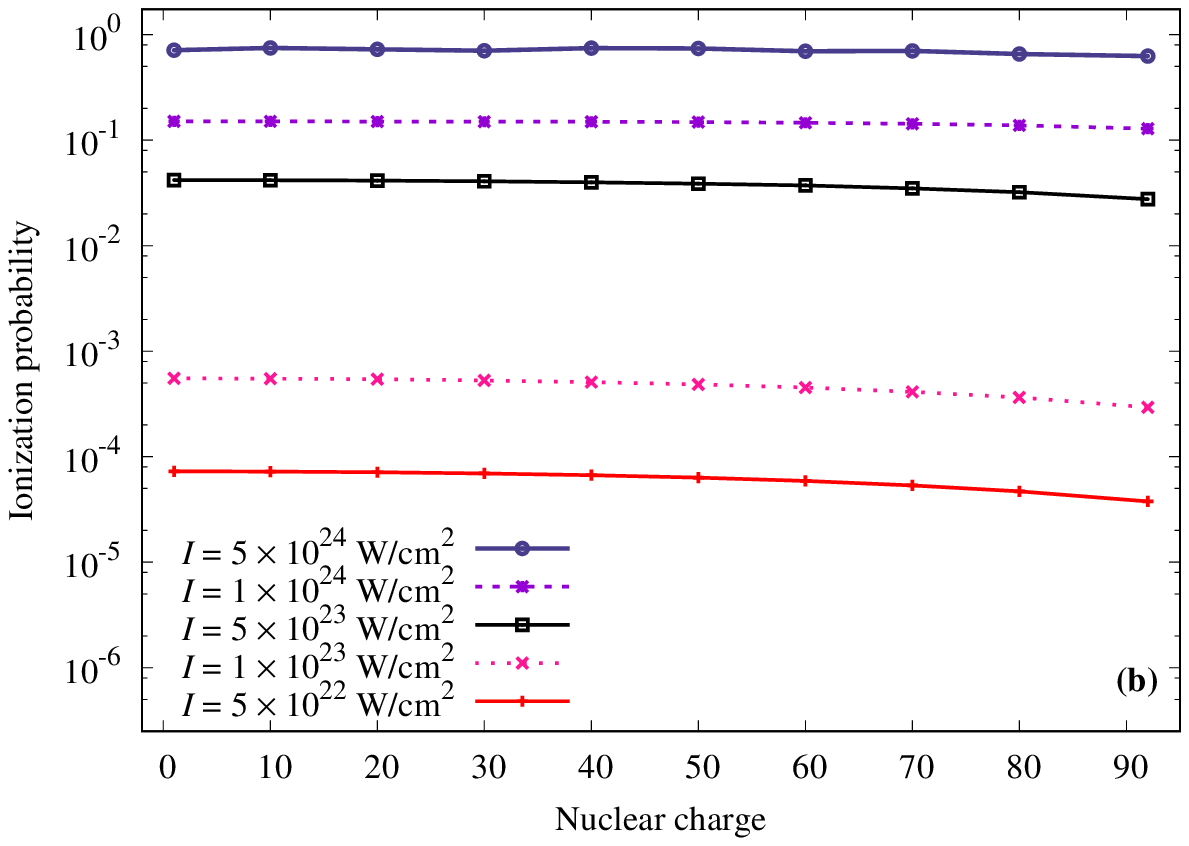}\\
\includegraphics[width=0.49\linewidth]{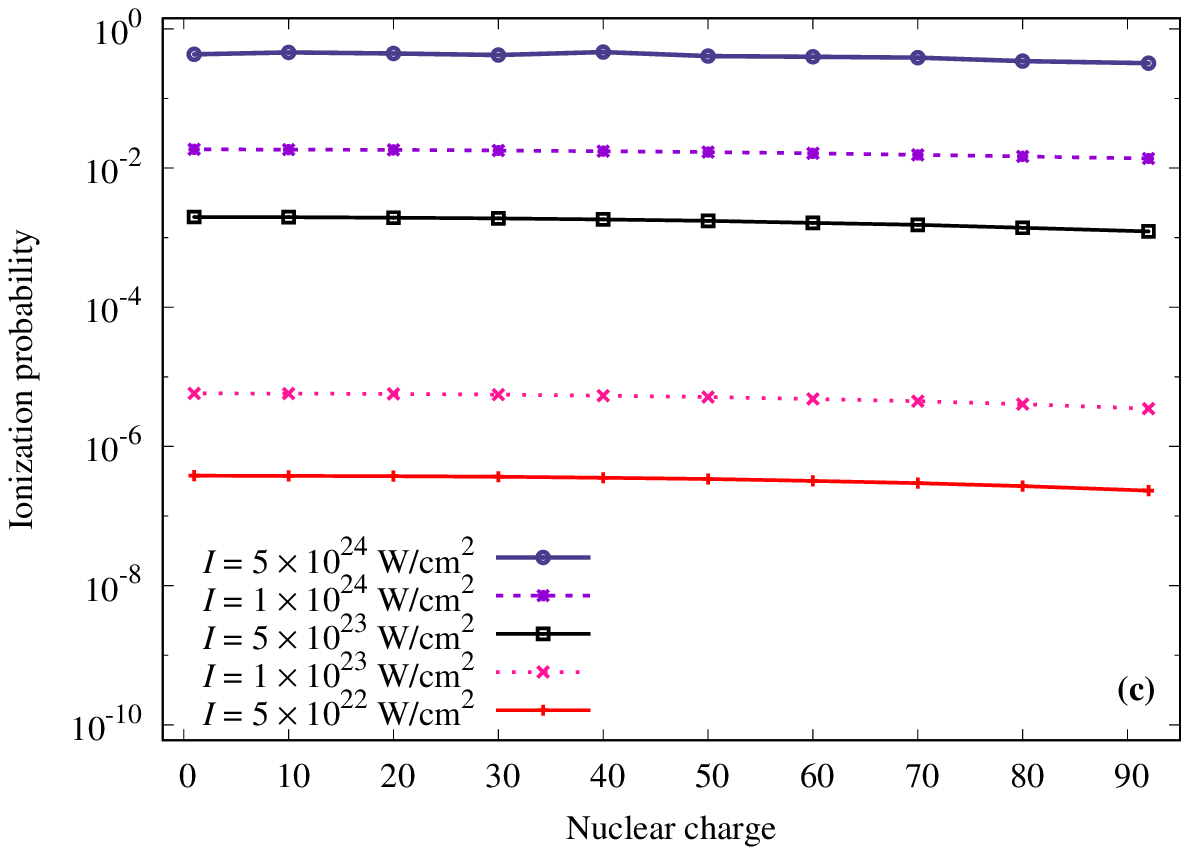}
\includegraphics[width=0.49\linewidth]{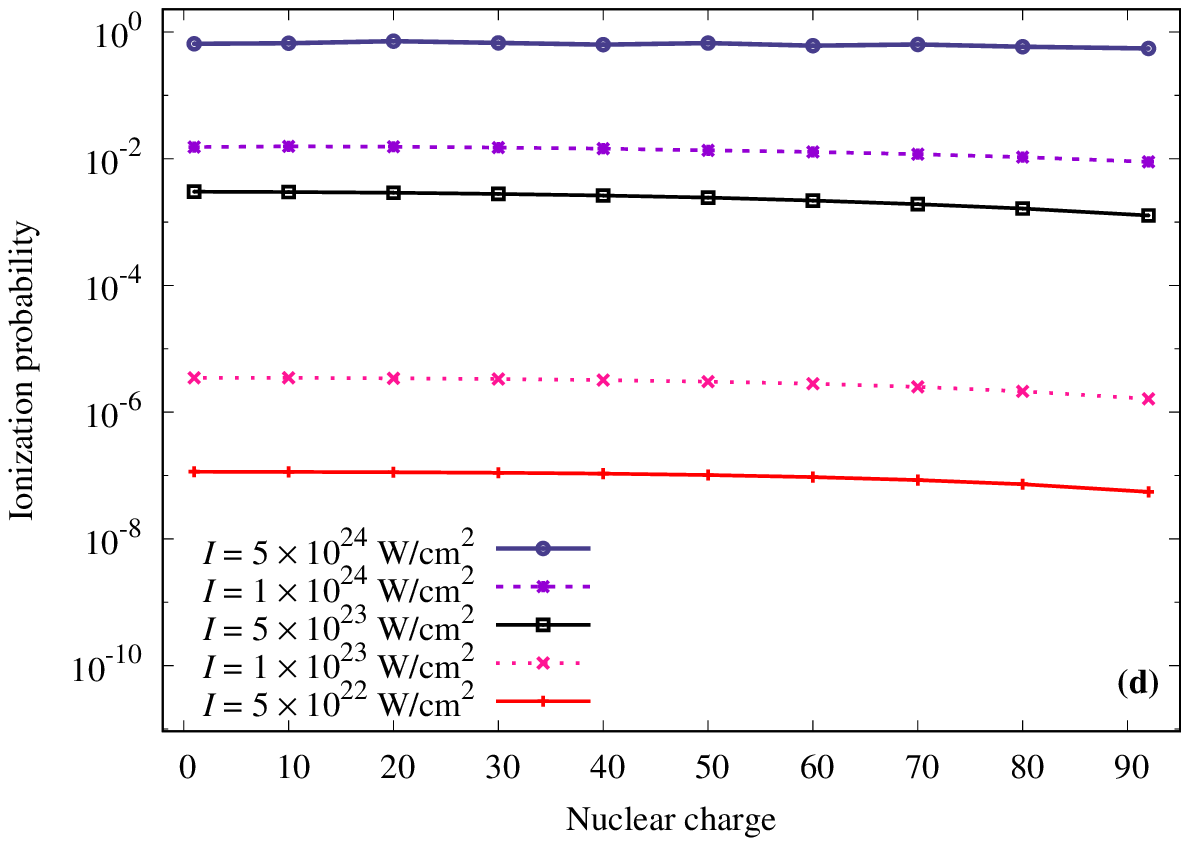}
\caption{Multiphoton ionization probability of the hydrogen-like ions as a function of the nuclear charge $Z$. (a) Two-photon ionization process with the laser wavelength of $0.06$~nm for $Z=50$. (b) Three-photon ionization process with the laser wavelength of $0.094$~nm for $Z=50$. (c) Four-photon ionization process with the laser wavelength of $0.12$~nm for $Z=50$. (d) Five-photon ionization process with the laser wavelength of $0.158$~nm for $Z=50$. In all subfigures, for $Z=50$, the peak intensity range is $5 \times 10^{22}$ to $5 \times 10^{24}$ W/cm$^2$. For the other ions, the laser peak intensity and wavelength are scaled according to Eq.~\eqref{eq:scaling2}.}
\label{fig:intensity}
\end{figure*}
\begin{figure*}[tb]
\includegraphics[width=0.49\linewidth]{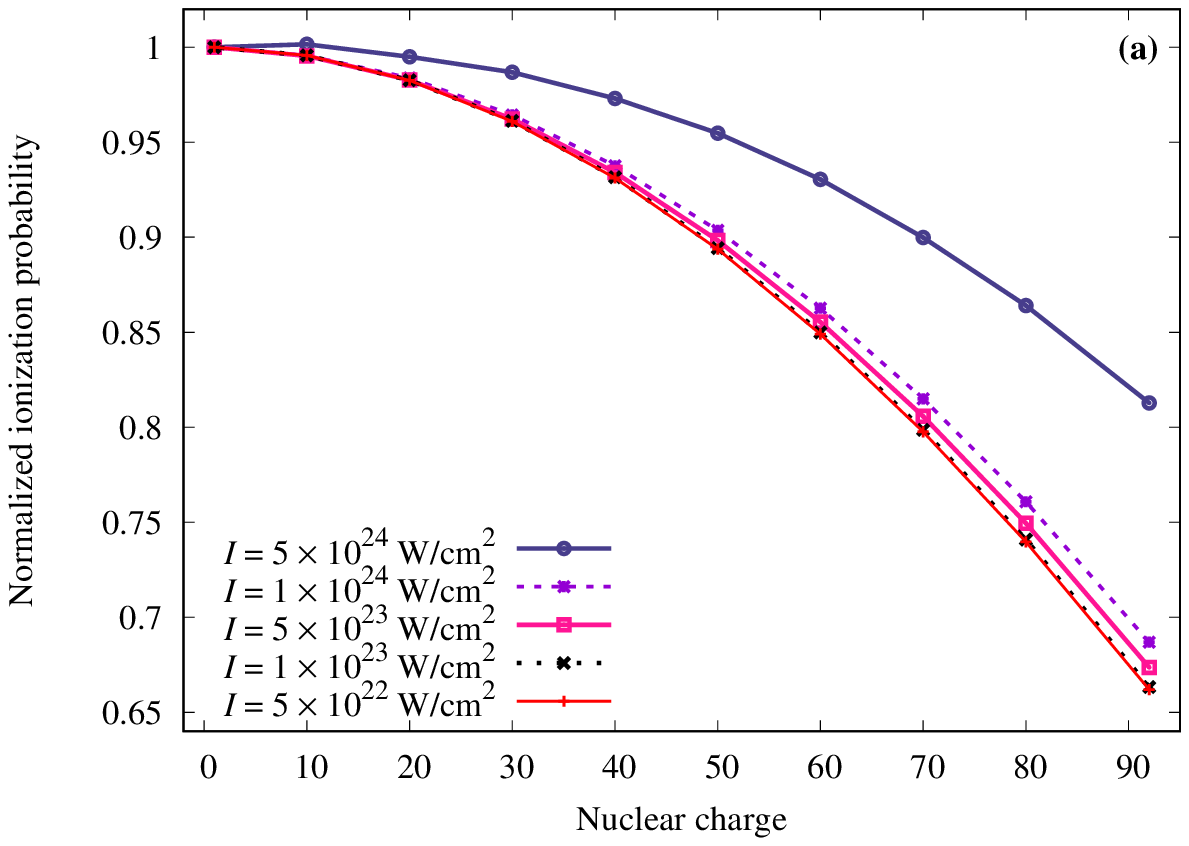}
\includegraphics[width=0.49\linewidth]{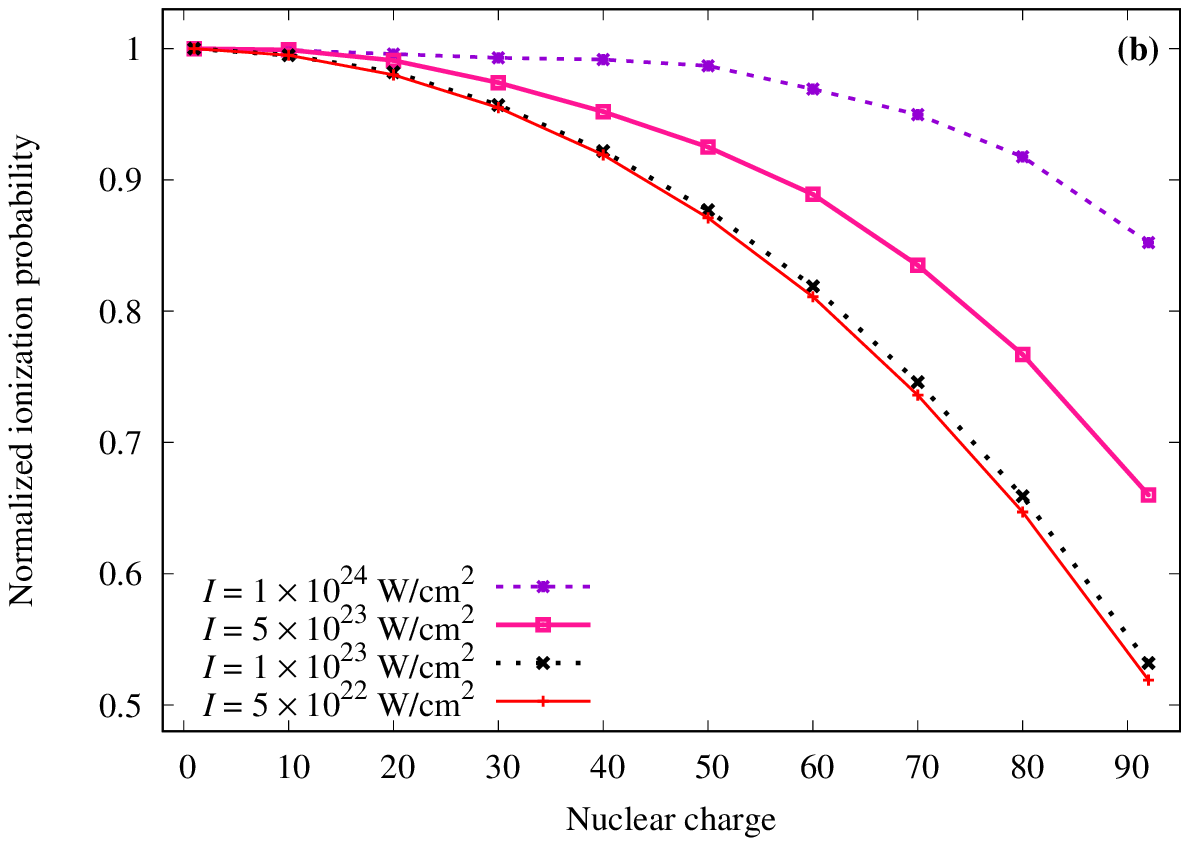}\\
\includegraphics[width=0.49\linewidth]{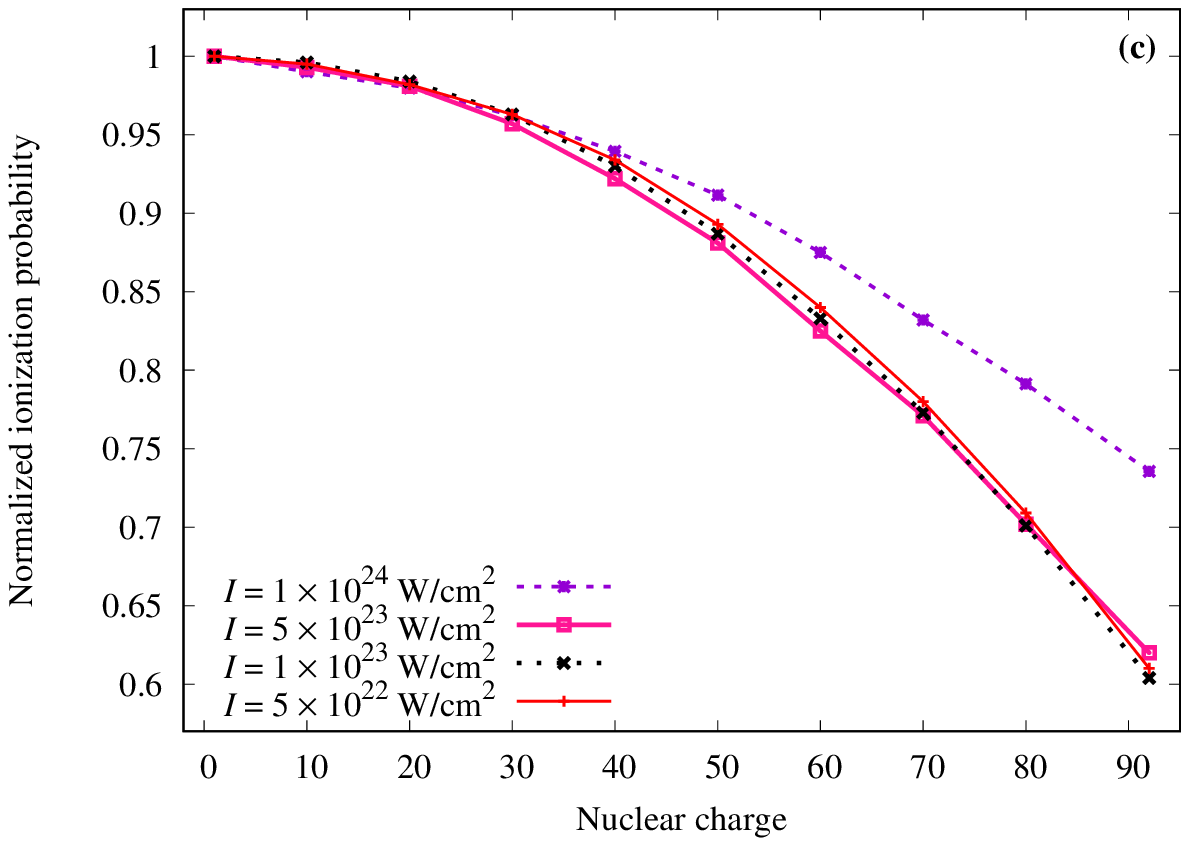}
\includegraphics[width=0.49\linewidth]{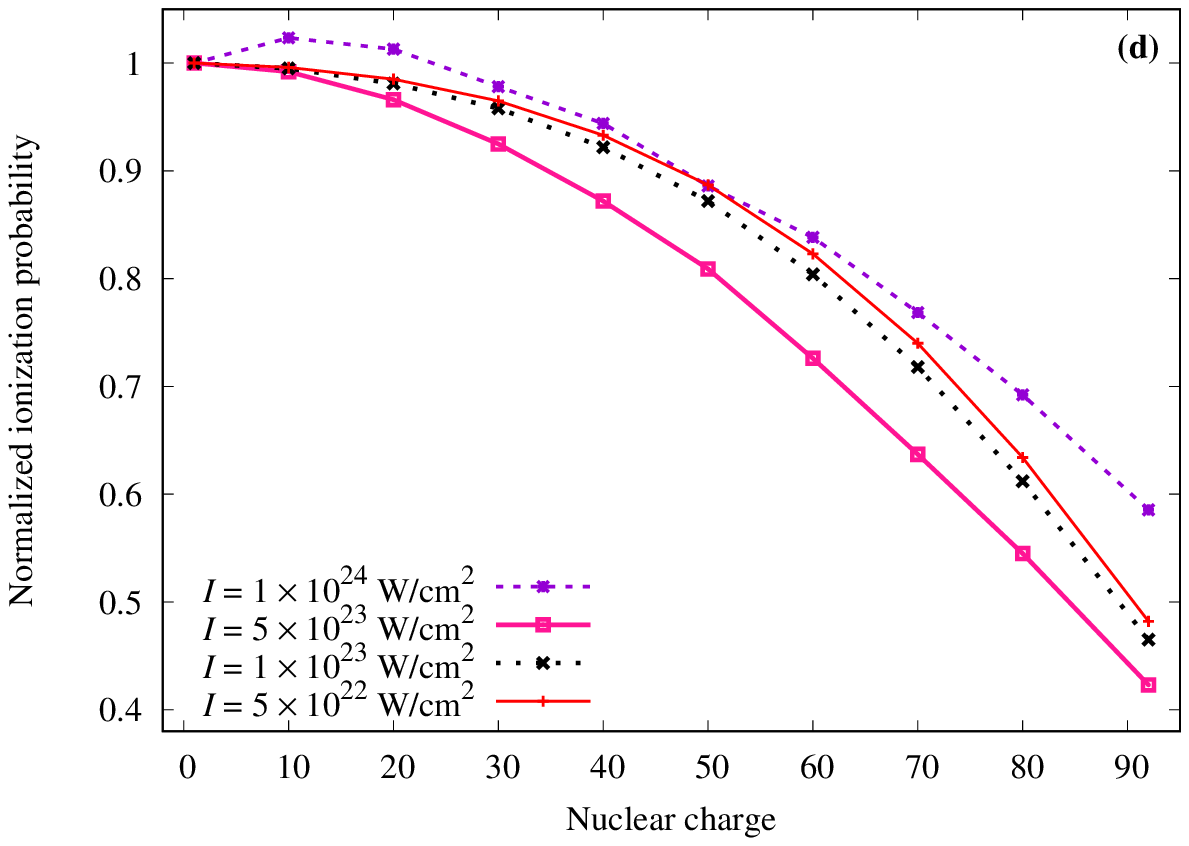}
\caption{Ionization probabilities of the hydrogen-like 
ions normalized to unity at $Z=1$. (a) For $Z=50$, the wavelength is $0.06$~nm (two-photon ionization), and the peak intensity range is $5 \times 10^{22}$ to $5 \times 10^{24}$~W/cm$^2$. (b) For $Z=50$, the wavelength is $0.094$~nm (three-photon ionization), and the peak intensity range is $5 \times 10^{22}$ to $1 \times 10^{24}$~W/cm$^2$. (c) For $Z=50$, the wavelength is $0.12$~nm (four-photon ionization), and the peak intensity range is $5 \times 10^{22}$ to $1 \times 10^{24}$~W/cm$^2$. (d) For $Z=50$, the wavelength is $0.158$~nm (five-photon ionization), and the peak intensity range is $5 \times 10^{22}$ to $1 \times 10^{24}$~W/cm$^2$. In all subfigures, for the other ions, the laser peak intensity and wavelength are scaled according to Eq.~\eqref{eq:scaling2}.}
\label{fig:rel_eff}
\end{figure*}

In Fig.~\ref{fig:New_TDDE_scaling}, we show the multiphoton 
ionization probabilities of several hydrogen-like ions for the same laser pulse 
parameters at $Z=50$ as in Fig.~\ref{fig:Z2_TDDE_scaling}. However, for the 
other nuclear charges the laser pulse parameters are calculated by the 
expressions~\eqref{eq:scaling2}. Compared with Fig.~\ref{fig:Z2_TDDE_scaling}, 
we have also extended the wavelength range to include the five-photon 
ionization process.

The new scaling relations used in 
Fig.~\ref{fig:New_TDDE_scaling} take into account the dominant relativistic 
effect, the lowering of the ground state energy level, that affects the multiphoton 
ionization process. Making use of these scaling laws allows us to nearly 
eliminate the shifts of the ionization curves in Fig.~\ref{fig:Z2_TDDE_scaling}. 
Looking at the figures~\ref{fig:Z2_TDDE_scaling} and~\ref{fig:New_TDDE_scaling}, 
one can notice that the correction of the (ground-state) ionization potential with the 
help of Eq.~\eqref{eq:scaling2} can change the ionization probabilities by 
orders of magnitude at some wavelengths (near the ionization thresholds). 
Therefore the combined scaling relations suggested in our work can be very useful for 
accurate predictions of the ionization dynamics of the hydrogen-like ions in a wide 
range of nuclear charges. The four curves in Fig.~\ref{fig:New_TDDE_scaling} 
are quite close to each other, but small discrepancies still exist. These 
deviations are caused by other relativistic effects not taken 
into account in Eq.~\eqref{eq:scaling2}.

\subsection{Intensity scaling} \label{ss:intensity_scaling}

In this section, we consider the scaling properties of the multiphoton ionization 
in a wide range of laser peak intensities. We have 
calculated the multiphoton ionization probability as a function of the  
nuclear charge $Z$ for five fixed peak intensities of the laser 
pulse~\eqref{eq:pulse}. For $Z=50$, we use the peak intensities of $5 \times 
10^{22}$, $1 \times 10^{23}$, $5 \times 10^{23}$, $1 \times 10^{24}$, and $5 
\times 10^{24}$ W/cm$^2$. For the other nuclear charges, the peak intensities 
are scaled as suggested by the expression~\eqref{eq:scaling2}. Two-, three-, 
four-, and five-photon ionization processes have been investigated (see 
Fig.~\ref{fig:intensity}). Here, we study 
nonresonant ionization, so the wavelengths 
have been chosen accordingly to avoid situations where the resonantly 
enhanced multiphoton ionization (REMPI) can take place.

In Fig.~\ref{fig:intensity}, one can see that the ionization probability is 
almost independent of the nuclear charge $Z$ for the range of intensities and 
multiphoton processes (from two-photon to five-photon ionization) under 
consideration. The results presented in Fig.~\ref{fig:intensity} cover a wide range of nuclear charges, laser wavelengths and pulse peak intensities. We can conclude that the scaling relations \eqref{eq:scaling2} are quite accurate and properly account for the dominant relativistic effect in multiphoton ionization. Below the remaining relativistic effects are discussed that cause the small deviations of the curves in Fig.~\ref{fig:intensity} from straight horizontal lines.

\subsection{Relativistic effects} \label{ss:rel_effects}

In the relativistic ionization regime discussed in the previous subsection, 
the ionization probabilities of the highly charged ions differ only slightly 
from the ionization probabilities of the hydrogen atom, if the laser pulse 
parameters are scaled using the new scaling laws \eqref{eq:scaling2}. The 
remaining small deviations of the curves in Fig.~\ref{fig:intensity} from the straight horizontal lines can be attributed to smaller relativistic effects not reflected in Eq.~\eqref{eq:scaling2}.

To further examine these small relativistic effects, we 
discuss here the ionization probabilities normalized to unity at $Z=1$. Displayed 
on a linear probability scale in Fig.~\ref{fig:rel_eff} are the same processes of two-, three-, four-, and five-photon ionization as shown on a logarithmic scale in 
Fig.~\ref{fig:intensity}. Two conclusions can be made when looking at Fig.~\ref{fig:rel_eff}. First, for each peak intensity, the normalized 
ionization probability decreases with increasing nuclear charge $Z$. That 
means, the scaling relations \eqref{eq:scaling2} overestimate the ionization 
probability of highly charged ions, and this effect becomes larger for larger 
$Z$. This is not surprising, since it is expected that any relativistic effects 
are more pronounced for heavier ions. In the range of the wavelengths and 
peak intensities used in the calculations, the dependence of the normalized 
probability on $Z$ is approximately quadratic. At the highest intensity 
$I = 5 \times 10^{24}$ W/cm$^2$, the saturation of the ionization 
is reached in the three-, four-, and five-photon ionization processes for most of 
the $Z$ values used in the calculations (not shown in Figs.~\ref{fig:rel_eff} (b), (c), (d)) and the quadratic dependence on $Z$ of the remaining relativistic effects breaks down. The saturation effects are visible even at the lower intensity $I = 1 \times 10^{24}$ W/cm$^2$ in Figs.~\ref{fig:rel_eff} (b) and (d). 

The second conclusion concerns the dependence of the normalized ionization probability on the peak intensity of the laser pulse when the saturation is not yet reached. Multiphoton ionization is an extremely nonlinear process, and its dependence on the intensity at each value of $Z$ can be non-monotonous and very complex, as one can see in Figs.~\ref{fig:rel_eff} (c) and (d).

The normalized ionization probabilities presented in the 
Fig.~\ref{fig:rel_eff} can help to isolate the small remaining relativistic effects, which still show up after compensation of the main 
relativistic effect due to the shift of the ionization potential by the scaling 
laws~\eqref{eq:scaling2}. For the multiphoton ionization processes studied here, 
the relativistic effects can be easily quantified and do not exceed 40\% of 
the nonrelativistic probabilities for the hydrogen atom (see 
Fig.~\ref{fig:rel_eff}). Such effects as, e.\,g.,  
spin-orbit coupling are quite small compared to the main relativistic 
effect, where the latter can cause a change of the ionization probabilities 
for up to four orders of magnitude even at the lowest intensity $I = 5 \times 10^{22}$ W/cm$^2$ 
used in the calculations (for example, compare the data for $Z=92$ in 
Figs.~\ref{fig:Z2_TDDE_scaling} and \ref{fig:New_TDDE_scaling}).
\section{Conclusion}  \label{s:conclusion}
In this paper, approximate scaling relations with respect to the nuclear charge 
have been presented for the TDDE describing hydrogen-like 
ions subject to laser fields within the dipole approximation. As a 
case study, the scalings relations~\eqref{eq:scaling2} have 
been applied to the multiphoton ionization yields of hydrogen-like ions 
with various nuclear charges in the two- to five-photon regime. The 
ion yields were calculated by solving the TDDE 
and the results obtained for different wavelengths and laser peak 
intensities, both scaled accordingly. While the previously derived 
non-relativistic scaling relations are found to be clearly insufficient 
in the relativistic regime described by the TDDE, the new scaling 
relations lead to good agreement between the ionization probabilities 
of the hydrogen-like ions with different nuclear charges. Noteworthy, 
depending on the nuclear charge and laser parameters, the new scaling factors 
modify not only the unscaled ion yields, but also the ones scaled by the 
non-relativistic scaling factor by several orders of magnitude and are 
thus very important even for order-of-magnitude estimates.

Also the dependence of the multiphoton ionization yields on the nuclear 
charge $Z$ for a wide range of the laser peak intensities has been investigated. 
It was found that the non-resonant two-, three-, four-, and five-photon 
ionization probabilities are almost $Z$-independent, if the laser parameters 
are scaled by the scaling laws \eqref{eq:scaling2} introduced in this work. 
This uniform behavior of the properly scaled results that covers a wide range 
of nuclear charges, laser wavelengths, and intensities is expected to be 
very useful for the planning and analysis of future experiments. 
Furthermore, the scaling relations may allow for a simple estimate of ion 
yields in laser fields of very high intensities, as they may be needed 
in corresponding laser-field induced plasma simulations or for considering 
possible radiation damage.  

The remaining small deviations of the scaled solutions of the TDDE reveal on the other hand the existence of further relativistic effects that are neither reflected in the non-relativistic scaling relations 
(as they are exact) nor in the relativistic shift of the ionization potential. 
Away from saturation these small relativistic effects not captured by the 
scaling relations proposed in this work are found to show an 
almost quadratic dependence on the nuclear charge.  

Finally, hydrogen-like ions with variable charge 
may be used as a tool for laser-pulse characterization or calibration, 
especially for light sources with extreme peak intensities. If the scaling 
relations are valid, the ion yield obtained with a laser pulse of, e.\,g., 
unknown laser peak intensity may be compared to the properly scaled ion 
yield obtained for an ion with lower nuclear charge exposed to a well 
characterized laser pulse of lower intensity. On the other hand, such 
comparisons could be used to uniquely identify experimentally beyond-dipole 
or other relativistic effects not yet contained in the scaling relations. This would be helpful for guiding subsequent theoretical studies. 
%
\section*{Acknowledgements}
This work was supported by RFBR (Grant No. 16-02-00233) and by SPbSU-DFG 
(Grants No. 11.65.41.2017 and No. STO 346/5-1). I.\,V.\,I.\ acknowledges the  support from the FAIR-Russia Research Center and the German-Russian Interdisciplinary Science Center (G-RISC) funded by the German Federal Foreign Office via the German Academic Exchange Service (DAAD). A.\,S.\ and I.\,V.\,I.\ acknowledge financial support by the German Ministry of Education and Research (BMBF) within APPA R\&D (Grant No.~05P15KHFAA and No.~05P18KHFAA). 


\begin{thebibliography}{44}
\bibitem{eli1} J.-P. Chambaret \textit{et al.}, Proceedings of SPIE \textbf{7721}, Solid State Lasers and Amplifiers IV, and High-Power Lasers, 77211D (2010). 

\bibitem{eli2} G. A. Mourou \textit{et al.}, \textit{ELI-Extreme Light Infrastructure Science and Technology with Ultra-Intense Lasers WHITEBOOK}
(THOSS Media GmbH, Berlin, 2011). 

\bibitem{xfel} T. Tschentscher and R. Feidenhans'l, Synchrotron Radiat. News \textbf{30}, 21 (2017).

\bibitem{lcls} M. Dunne and B. Schoenlein, Synchrotron Radiat. News \textbf{30}, 7 (2017).


\bibitem{vogel} M. Vogel, W. Quint, G. G. Paulus, and Th. St$\mathrm{\ddot{o}}$hlker, Nucl. Instrum. Methods B \textbf{285}, 65 (2012).

\bibitem{ringleb} S. Ringleb, M. Vogel, S. Kumar, W. Quint, G. G. Paulus, and Th. St$\mathrm{\ddot{o}}$hlker, Phys. Scr. \textbf{T166}, 014067 (2015).

\bibitem{phelix} PHELIX -- The Petawatt High-Energy Laser for Heavy Ion EXperiments, 2018 (accessed June 27, 2018), \url{https://www.gsi.de/en/work/research/appamml/plasma_physicsphelix/phelix.htm}

\bibitem{selsto} S. Selst{\o}, E. Lindroth, and J. Bengtsson, Phys. Rev. A \textbf{79}, 043418 (2009).

\bibitem{pindzola} M. S. Pindzola, J. A. Ludlow, and J. Colgan, Phys. Rev. A \textbf{81}, 063431 (2010).

\bibitem{pindzola2} M. S. Pindzola, Sh. A. Abdel-Naby, F. Robicheaux, and J. Colgan, Phys. Rev. A \textbf{85}, 032701 (2012).

\bibitem{vanne} Y. V. Vanne and A. Saenz, Phys. Rev. A \textbf{85}, 033411 (2012).

\bibitem{rozenbaum} E. B. Rozenbaum, D. A. Glazov, V. M. Shabaev, K. E. Sosnova, and D. A. Telnov, Phys. Rev. A \textbf{89}, 012514 (2014).

\bibitem{klaiber} M. Klaiber and K. Z. Hatsagortsyan, Phys. Rev. A \textbf{90}, 063416 (2014).

\bibitem{klaiber2} M. Klaiber, E. Yakaboylu, C. M$\mathrm{\ddot{u}}$ller, H. Bauke, G. G. Paulus, K. Z. Hatsagortsyan, J. Phys. B: At. Mol. Opt. Phys. \textbf{47}, 065603 (2014).

\bibitem{ivanova} I. V. Ivanova, A. I. Bondarev, I. A. Maltsev, D. A. Tumakov, and V. M. Shabaev, J. Phys.: Conf. Ser. \textbf{635}, 092040 (2015).

\bibitem{simonsen} A. S. Simonsen, T. Kjellsson, M. F{\o}rre, E. Lindroth, and S. Selst{\o},
Phys. Rev. A \textbf{93}, 053411 (2016).

\bibitem{ivanova2} I. V. Ivanova, A. Saenz, A. I. Bondarev, I. A. Maltsev, V. M. Shabaev, D. A. Telnov, J. Phys.: Conf. Ser. \textbf{875}, 022031 (2017).

\bibitem{kjellsson} T. Kjellsson, S. Selst{\o}, and E. Lindroth, Phys. Rev. A \textbf{95}, 043403 (2017).

\bibitem{kjellsson2} T. Kjellsson, M. F{\o}rre, A. S. Simonsen, S. Selst{\o}, and E. Lindroth, Phys. Rev. A \textbf{96}, 023426 (2017).

\bibitem{telnov} D. A. Telnov, D. A. Krapivin, J. Heslar, and S.-I. Chu, J. Phys. Chem. A \textbf{122}, 8026 (2018).

\bibitem{adk} M. V. Ammosov, N. B. Delone, and V. P. Krainov, JETP \textbf{64}, 1191 (1986).

\bibitem{lambropoulos} P. Lambropoulos and X. Tang, JOSAB \textbf{4}, 821 (1987).

\bibitem{madsen} L. B. Madsen and P. Lambropoulos, Phys. Rev. A \textbf{59}, 4574 (1999).

\bibitem{alnaser} A. S. Alnaser, X. M. Tong, T. Osipov, S. Voss, C. M. Maharjan, B. Shan, Z. Chang, and C. L. Cocke, Phys. Rev. A \textbf{70}, 023413 (2004).

\bibitem{pullen} M. G. Pullen, W. C. Wallace, D. E. Laban, A. J. Palmer, G. F. Hanne, A. N. Grum-Grzhimailo, K.~Bartschat, I. Ivanov, A. Kheifets, D. Wells, H. M. Quiney, X. M. Tong, I. V. Litvinyuk, R.~T.~Sang, and D. Kielpinski,
Phys. Rev. A \textbf{87}, 053411 (2013).

\bibitem{zhao} S.-F. Zhao, A.-T. Le, C. Jin, X. Wang, and C. D. Lin, Phys. Rev. A \textbf{93}, 023413 (2016).

\bibitem{wallace} W. C. Wallace, O. Ghafur, C. Khurmi, Satya Sainadh U, J. E. Calvert, D. E. Laban, M. G. Pullen, K.~Bartschat, A. N. Grum-Grzhimailo, D. Wells, H. M. Quiney, X. M. Tong, I. V. Litvinyuk, R.~T.~Sang, and D. Kielpinski,
Phys. Rev. Lett. \textbf{117}, 053001 (2016).

\bibitem{deboor} C. de Boor, \textit{A Practical Guide to Splines}, Applied Mathematical Sciences, Vol. 27, revised edition (Springer-Verlag, New York, 2001).

\bibitem{johnson} W. R. Johnson, S. A. Blundell, and J. Sapirstein, Phys. Rev. A \textbf{37}, 307 (1988).

\bibitem{shabaev} V. M. Shabaev, I. I. Tupitsyn, V. A. Yerokhin, G. Plunien, and G. Soff, Phys. Rev. Lett. \textbf{93}, 130405 (2004).

\bibitem{crank} J. Crank and P. Nicolson, Proc. Cambridge Philos. Soc. \textbf{43}, 50 (1947).

\bibitem{fleck} J. A. Fleck, J. R. Morris, and M. D. Feit, Appl. Phys. \textbf{10}, 129 (1976).

\bibitem{press} W. H. Press, S. A. Teukolsky, W. T. Vetterling ,and B. P. Flannery, \textit{Numerical Recipes in Fortran 77}, 2nd ed., Vol. 1 of Fortran
Numerical Recipes (Cambridge University Press, Cambridge, 1992).

\bibitem{varshalovich}   D. A. Varshalovich, A. N. Moskalev, and V. K. Khersonskii, \textit{Quantum Theory of Angular Momentum} (World Scientific, Singapore, 1988).

\end{thebibliography}
\end{document}